\documentclass[twocolumn,table,xcdraw]{aastex63}

\definecolor{auburn}{rgb}{0.43, 0.21, 0.1}
\definecolor{bostonuniversityred}{rgb}{0.8, 0.0, 0.0}
\definecolor{royalazure}{rgb}{0.0, 0.22, 0.66}
\definecolor{shamrockgreen}{rgb}{0.0, 0.62, 0.38}
\definecolor{myBlue}{HTML}{5480f1}
 \hypersetup{linkcolor=royalazure,citecolor=royalazure,filecolor=cyan,urlcolor=myBlue}
\usepackage[normalem]{ulem}

\usepackage{physics}

\newcommand\PdV{P\mathrm{d}V}
\newcommand\divv{\nabla \cdot \vb*{v}}

\usepackage{lipsum}  
\usepackage{amsmath}

\received{date}
\revised{date}
\accepted{date}
\submitjournal{ApJ}

\shorttitle{Warping away gravitational instabilities}
\shortauthors{Rowther, Nealon, \& Meru}
\graphicspath{{./}{Images/}}
\begin{document}

\title{Warping Away Gravitational Instabilities in Protoplanetary Discs}

\correspondingauthor{Sahl Rowther}
\email{sahl.rowther@warwick.ac.uk}

%\correspondingauthor{Farzana Meru}
%\email{f.meru@warwick.ac.uk}

\author[0000-0003-4249-4478]{Sahl Rowther}
\affiliation{Centre for Exoplanets and Habitability, University of Warwick, Coventry CV4 7AL, UK}
\affiliation{Department of Physics, University of Warwick, Coventry CV4 7AL, UK}

\author[0000-0003-0856-679X]{Rebecca Nealon}
\affiliation{Centre for Exoplanets and Habitability, University of Warwick, Coventry CV4 7AL, UK}
\affiliation{Department of Physics, University of Warwick, Coventry CV4 7AL, UK}
\author[0000-0002-3984-9496]{Farzana Meru}

\affiliation{Centre for Exoplanets and Habitability, University of Warwick, Coventry CV4 7AL, UK}
\affiliation{Department of Physics, University of Warwick, Coventry CV4 7AL, UK}
%




\begin{abstract}
	We perform 3D SPH simulations of warped, non-coplanar gravitationally unstable discs to show that as the warp propagates through the self-gravitating disc, it heats up the disc rendering it gravitationally stable. Thus losing their spiral structure and appearing completely axisymmetric. In their youth, protoplanetary discs are expected to be massive and self-gravitating, which results in non-axisymmetric spiral structures. However recent observations of young protoplanetary discs with ALMA have revealed that discs with large-scale spiral structure are rarely observed in the midplane. Instead, axisymmetric discs with some also having ring \& gap structures are more commonly observed. Our work invloving warps, non-coplanar disc structures that are expected to commonly occur in young discs, potentially resolves this discrepancy between observations and theoretical predictions. We demonstrate that they are able to suppress the large-scale spiral structure of self-gravitating protoplanetary discs.
\end{abstract}



\section{Introduction} \label{sec:intro}

In recent years, a large number of protoplanetary discs have been observed at millimeter wavelengths with ALMA. Most of these discs are axisymmetric, with some also containing rings and gaps \citep{2015ALMA,2016Andrews,2018Fedele,2018Andrews,2018Huang,2018Dipierro,2020Booth}, even though some of them are quite young ($<1$Myr). Ring and gap structures have also been observed in even younger ($\,\lesssim 0.5\,$Myrs) Class 1 discs \citep{2018Sheehan,Segura-Cox2020}. Young discs are thought to be massive and could potentially be gravitationally unstable. A characteristic feature of such discs are large scale spiral features. There is evidence that discs with spiral arms in the midplane exist \citep{2016Perez, 2018bHuang}, although they seem to be quite rare.

Does the lack of observed large scale spiral structures imply that young discs are not as massive as expected? Or can signatures of gravitational instabilities be hidden? It is reasonable to assume that gravitationally unstable discs do not evolve in isolation. The physical processes that are often used to explain observed substructures such as rings and gaps, along with the mechanisms that can warp a disc will also influence the evolution of young self-gravitating discs. A common explanation for rings and gaps are planet-disc interactions which can also suppress spiral structures in gravitationally unstable discs \citep{2020bRowther}. 

There are several mechanisms that can warp a protoplanetary disc. These include a misaligned internal (planetary or stellar) companion, a misaligned unbound stellar companion (flyby), or misaligned infalling material from chaotic accretion episodes. The rate at which these discs are observed also suggests that warps and misalignments are common in protoplanetary discs \citep{2017Benisty,2017Walsh,2018Casassus,2021Ballabio}. The latter two mechanisms are more relevant for gravitationally unstable discs. Flybys are much more common earlier in a disc's lifetime and become less frequent over time \citep{2013Pfalzner,2016Vincke,2018Bate}. Misaligned infall is also expected to occur very early in a disc's lifetime when there is still plenty of material around the protostar \citep{2018Bate,2019Sakai}. Both of these mechanisms are more likely to occur precisely when discs are more likely to be gravitationally unstable and can alter the evolution of the disc.

% The evolution of a disc that has been warped can be very different to a flat disc. In the warped region of the disc, a horizontal pressure gradient which varies with $z$ is induced. The magnitude of the induced pressure gradient is given by \citep{2007Lodato}
% \begin{equation}
% 	\label{eq:pres}
% 	\frac{\partial p}{\partial R} \sim \frac{\partial p}{\partial z} \psi \sim \frac{p \psi}{H}.
% \end{equation}

There have been some simulations of flybys interacting with a gravitationally unstable disc, but in these studies the flybys were coplanar or head-on. Additionally, the distance of closest approach occurred inside the disc causing major disruption to the disc structure and completely suppressing gravitational instabilities \citep{2007bLodato,2009Forgan}. These works demonstrate that a flyby altering the disc structure can affect the heating and thus the self-gravitating structures. However, they do not consider warps or misalignments. \cite{2010Thies} on the other hand did study the impact of inclined flybys and found that stellar encounters induced fragmentation. However, their disc masses were much more massive and hence more prone to fragmentation.

In this paper we use three-dimensional global numerical simulations to consider the evolution of an isolated self-gravitating disc subjected to a warp. To our knowledge, this is the first study of a self-gravitating disc subjected to a warp. This paper is organised as follows. In \S\ref{sec:WT} we recall the relevant warp theory and its context in self-gravitating discs. In \S\ref{sec:model} we describe the simulations presented in this work. In \S\ref{sec:Res} we present our results of the impact that an artificially introduced warp has on the structure and evolution of gravitationally unstable protoplanetary discs. We discuss our work in the context of its limitations and observations in \S\ref{sec:disc}. We conclude our work in \S\ref{sec:conc}.

\begin{figure}
	\begin{center}
		\includegraphics[width=\linewidth]{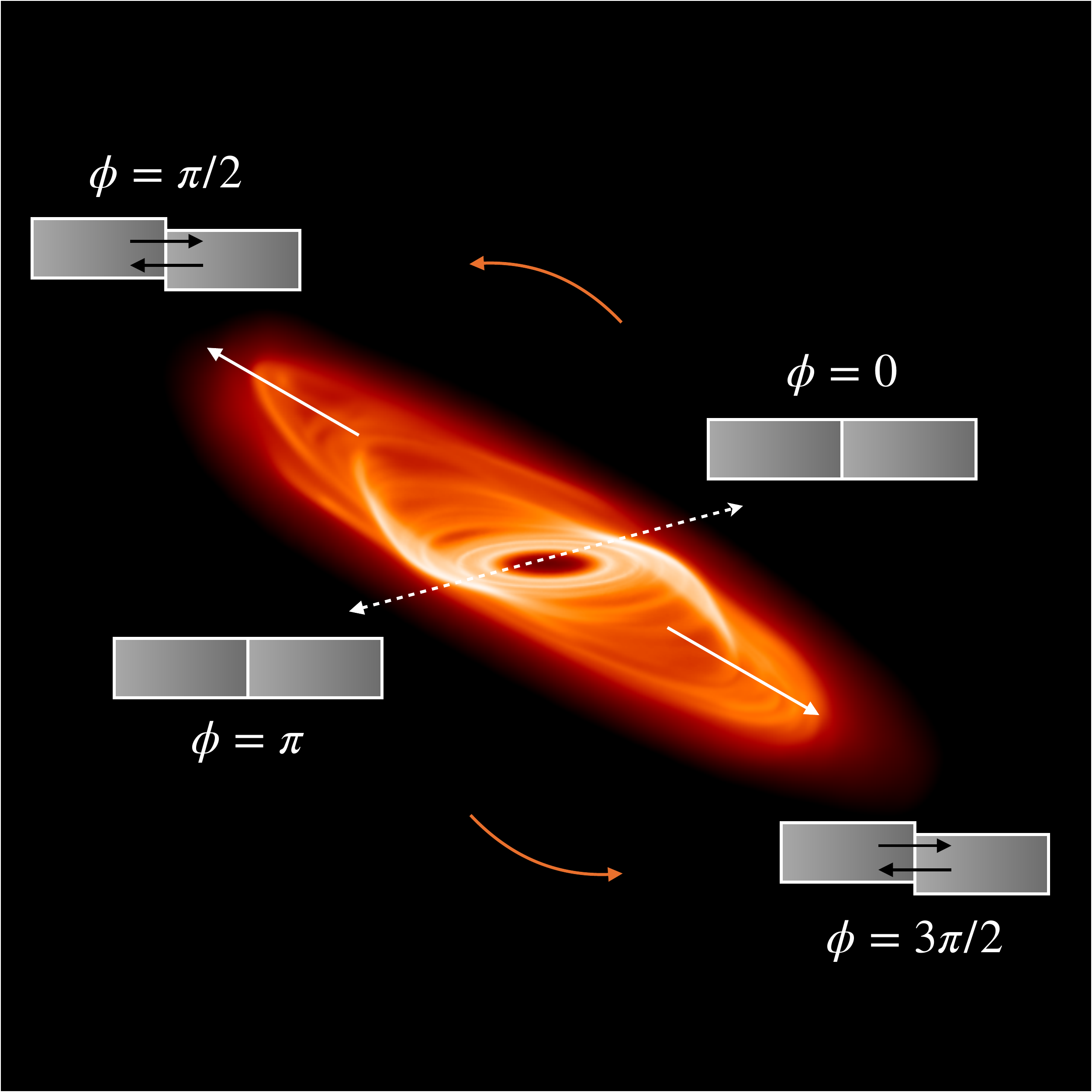}
		\caption{A 3D projection of the disc after it has been warped. The annotations \citep[see Fig 10 of][]{2007Lodato} show the radial pressure gradients induced by the warp at different azimuths as a fluid element orbits the disc. The orange arrows show the rotation of the disc. The shaded grey boxes represent the pressure between adjacent annuli of gas. At $\phi = 0$ and $\phi = \pi$ the annuli are aligned, hence there is no pressure gradient. At all other azimuths, the heights of the annuli are offset resulting in an oscillating radial pressure gradient as the gas traverses an orbit. The direction of the resulting pressure gradient is shown by the black arrows. The colour scale represents the surface density of the disc with brighter colours correspond to regions of higher density.}
		\label{initWarp}
	\end{center}
\end{figure}

\section{Warp Theory}
\label{sec:WT}

Warped discs are characterised by a radial dependence on the angular momentum vector. The steepness of the warp is described by the warp amplitude $\psi$ which is given by \citep[eg.][]{2010Lodato}

\begin{equation}
	\psi = R \left | \frac{\partial \vb*{l}}{\partial R} \right|,
\end{equation}
where $\vb*{l}$ is the unit angular momentum vector, and $R$ is the radial location spherical coordinates. 

To demonstrate the effect that the warp has on the disc, in Figure \ref{initWarp} we show a 3D projection of a warped gravitationally unstable disc with annotations to illustrate how the warp propagates through the disc. Consider a fluid element as it orbits around the disc with a position given by $\phi$. Here the disc may be described as concentric rings or annuli of gas and the warp causes adjacent rings to be vertically offset. The rotation of the disc is represented by the orange arrows. The shaded grey boxes represent the direction of the pressure between adjacent annuli of gas. At $\phi = 0\ \mathrm{and}\ \pi$, adjacent rings will be perfectly aligned with no variation in the vertical height $z$, hence at these locations there is no pressure gradient. At all other azimuths, adjacent rings will have variations in $z$, resulting in pressure gradients due to overpressure regions above or below the local midplane. The magnitude of the induced pressure gradient is given by \citep{2007Lodato}

\begin{equation}
	\label{eq:pres}
	\frac{\partial P}{\partial R} \sim \frac{\partial P}{\partial z} \psi \sim \frac{P \psi}{H},
\end{equation}
where $P$ is the pressure, $z$ is the vertical height, and $H$ is disc scale height. The direction of the resulting pressure gradient, represented by the black arrows, changes based on whether the fluid element is flowing towards or away from misaligned regions. As it moves towards vertically offset regions, away from $\phi = 0$ and $\pi$, the resulting radial pressure gradient points inwards. The direction of the pressure gradient is reversed when flowing away from the vertically offset regions, towards $\phi = 0$ and $\pi$. Hence the fluid element feels an oscillating pressure gradient as it orbits in the warp which will drive the disc's evolution.
This oscillating radial pressure gradient can trigger a strong response in the velocity flow in the disc. The response for example in the radial velocity field from \cite{2007Lodato}

\begin{equation}
	\label{eq:vr}
	v_{R} \propto \psi \cos\phi
\end{equation}
depends on the warp amplitude of the disc and the azimuthal angle $\phi$.

In typical studies of warped protoplanetary discs, the equation of state is assumed to be isothermal, where $\PdV$ work is neglected \citep{2010Lodato}. However, in the work presented here where we consider warps in gravitationally unstable discs, which contains large scale spiral structures that cause shocks, $\PdV$ work is naturally important. Hence, if a warp can excite strong responses in the velocity flow, the spiral structures in the disc are expected to be impacted due to additional $\PdV$ heating.

\section{Method} \label{sec:model}

\subsection{Hydrodynamical simulations \& initial conditions}

We use \textsc{Phantom}, a smoothed particle hydrodynamics (SPH) code developed by \cite{2018Price} to perform the suite of simulations presented here.

The disc is modelled using 2 million particles between $R_{\mathrm{in}} = 3$ and $R_{\mathrm{out}} = 150$ in code units with a disc-to-star mass ratio of 0.1. The central star is modelled using a fixed external potential.
The initial surface mass density is set as a smoothed power law and is given by

\begin{equation}
	\Sigma = \Sigma_{\mathrm{in}}  \left ( \frac{R}{R_{\mathrm{in}}} \right)^{-1} f_{s},
\end{equation}
where $\Sigma_{\mathrm{in}}$ is the surface mass density at $R=R_{\mathrm{in}}$ and ${f_{s} = 1-\sqrt{R_{\mathrm{in}}/R}}$ is {the factor used to smooth the surface density at the inner boundary of the disc}. The initial temperature profile is expressed as a power law

\begin{equation}
	T = T_{\mathrm{in}} \left ( \frac{R}{R_{\mathrm{in}}} \right)^{-0.5},
\end{equation}
where $T_{\mathrm{in}}$ is set such that the disc aspect ratio ${H/R=0.05}$ at $R=R_{\mathrm{in}}$. The internal energy equation is

\begin{equation}
    \label{eq:enrg}
	\frac{\mathrm{d}u}{\mathrm{d}t} = -\frac{P}{\rho} \left ( \divv \right) + \Lambda_{\mathrm{shock}} - \frac{\Lambda_{\mathrm{cool}}}{\rho}
\end{equation}
where we assume an adiabatic equation of state, and $u$ is the specific internal energy. The first term on the RHS is the $P\mathrm{d}V$ work, $\Lambda_{\mathrm{shock}}$ is a heating term that is due to the artificial viscosity used to correctly deal with shock fronts, and

\begin{equation}
	\Lambda_{\mathrm{cool}} = \frac{\rho u}{t_{\mathrm{cool}}}
\end{equation}
controls the cooling in the disc. Here, the cooling time is modelled using a simple prescription such that it is proportional to the dynamic time by a constant factor $\beta_{\mathrm{cool}}$ \citep{2001Gammie},

\begin{equation}
	t_{\mathrm{cool}} = \beta_{\mathrm{cool}}\,\Omega^{-1},
\end{equation}
where $\Omega$ is the orbital frequency. 
Assuming the transfer of angular momentum is locally driven by gravitoturbulence \citep{2001Gammie}, $\beta_{\mathrm{cool}}$ can be related to the $\alpha$ viscosity by \citep{1973SS}

\begin{equation}
    \alpha = \frac{4}{9}\frac{1}{\gamma(\gamma - 1)} \frac{1}{\beta_{\mathrm{cool}}}.
\end{equation}
Here $\beta_{\mathrm{cool}} = 15$, which using the above equation gives a theoretical $\alpha = 0.027$. To model shocks, we use an artificial viscosity switch that utilises the time derivative of the velocity divergence introduced by \cite{2010Cullen}. The artificial viscosity parameter $\alpha_{v}$ varies depending on the proximity to a shock. It takes a maximum of $\alpha_{\mathrm{max}} = 1$ close to the shock and a minimum of $\alpha_{\mathrm{min}} = 0$ far away. The artificial viscosity coefficient $\beta_{v}$ is set to 2 (see \citealt{2018Price,2015Nealon}).

\begin{figure*}
	\begin{center}
		\includegraphics[width=\linewidth]{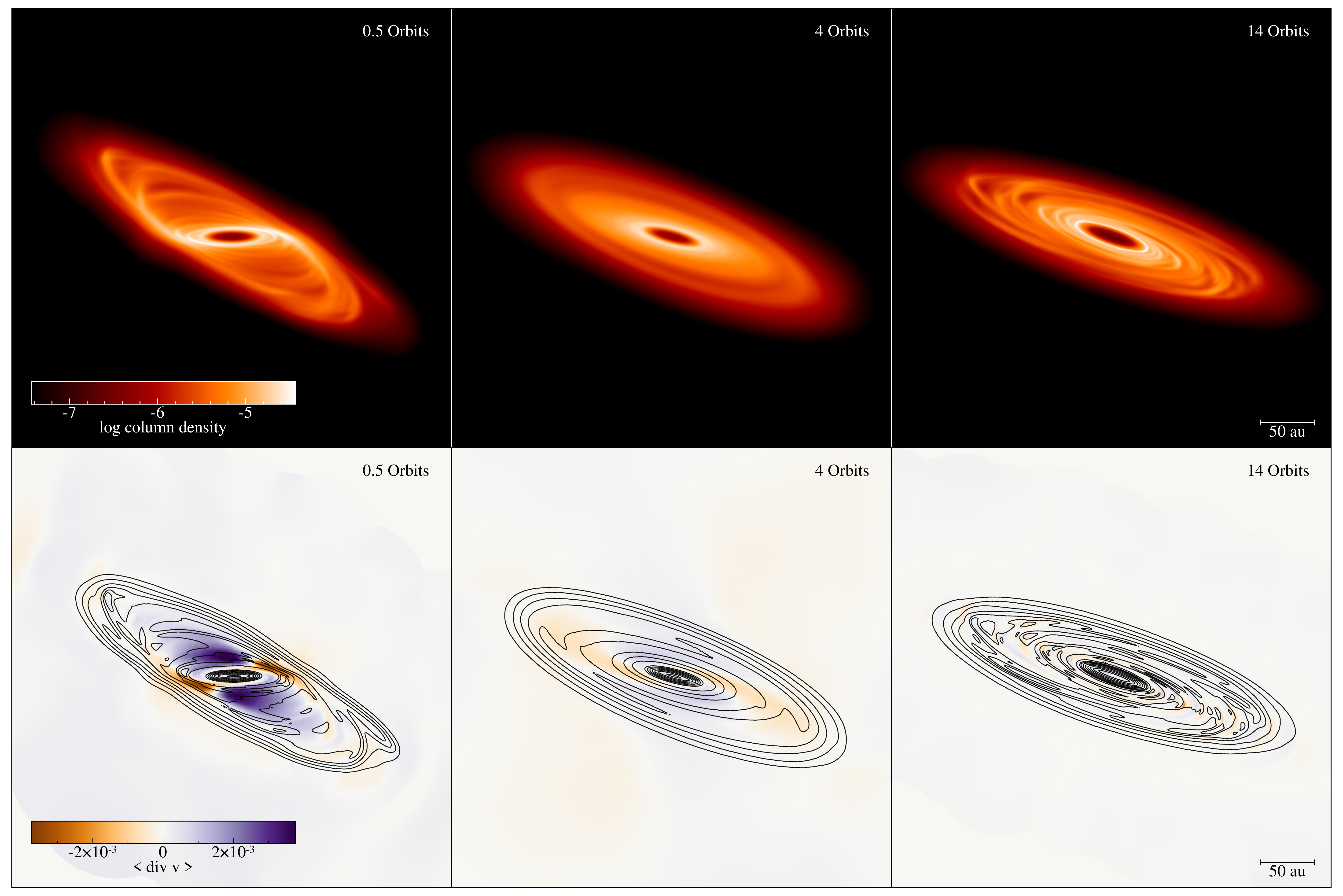}
		\caption{Plots of the surface density, $\Sigma$ (top) and divergence of velocity, $\divv$ (bottom) in code units where the angled brackets represent a density weighted average, showing the evolution of a $0.1M_{\odot}$ disc 0.5, 4, and 14 orbits (from left to right) after a warp is introduced. The warp induces a strong response in the velocity flow of the disc (left panels), which heats it up resulting in an axisymmetric gravitationally stable disc (middle panels). After the disc has realigned, cooling takes over with the disc eventually recovering its spiral features (right panels).}
		\label{Q_R75i30}
	\end{center}
\end{figure*}

\begin{figure}
	\begin{center}
		\includegraphics[width=\linewidth]{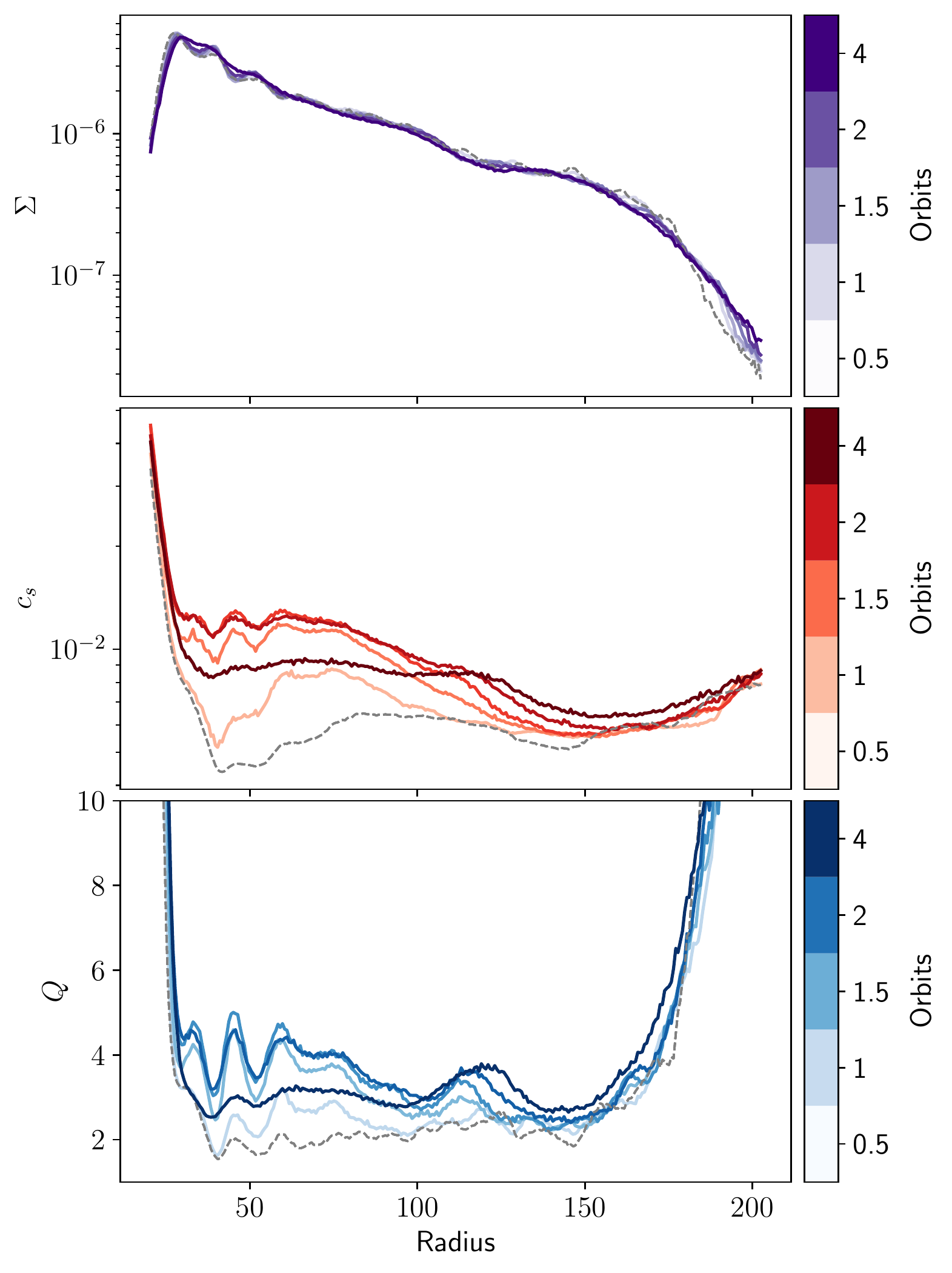}
		\caption{Azimuthally averaged surface density, $\Sigma$ (top), the sound speed, $c_{s}$ (middle), and the Toomre parameter, $Q$ (top) in code units at 0.5, 1, 1.5, 2, and 4 orbits after the warp has been introduced. The darker shades represent later times. The dashed gray line in each panel is the initial profile at the moment the warp is introduced. The increase in $c_{s}$ due to the warp is the reason for the increase in $Q$ causing the disc to become gravitationally stable. }
		\label{Q_R75i30azi}
	\end{center}
\end{figure}

\begin{figure}
	\begin{center}
		\includegraphics[width=\linewidth]{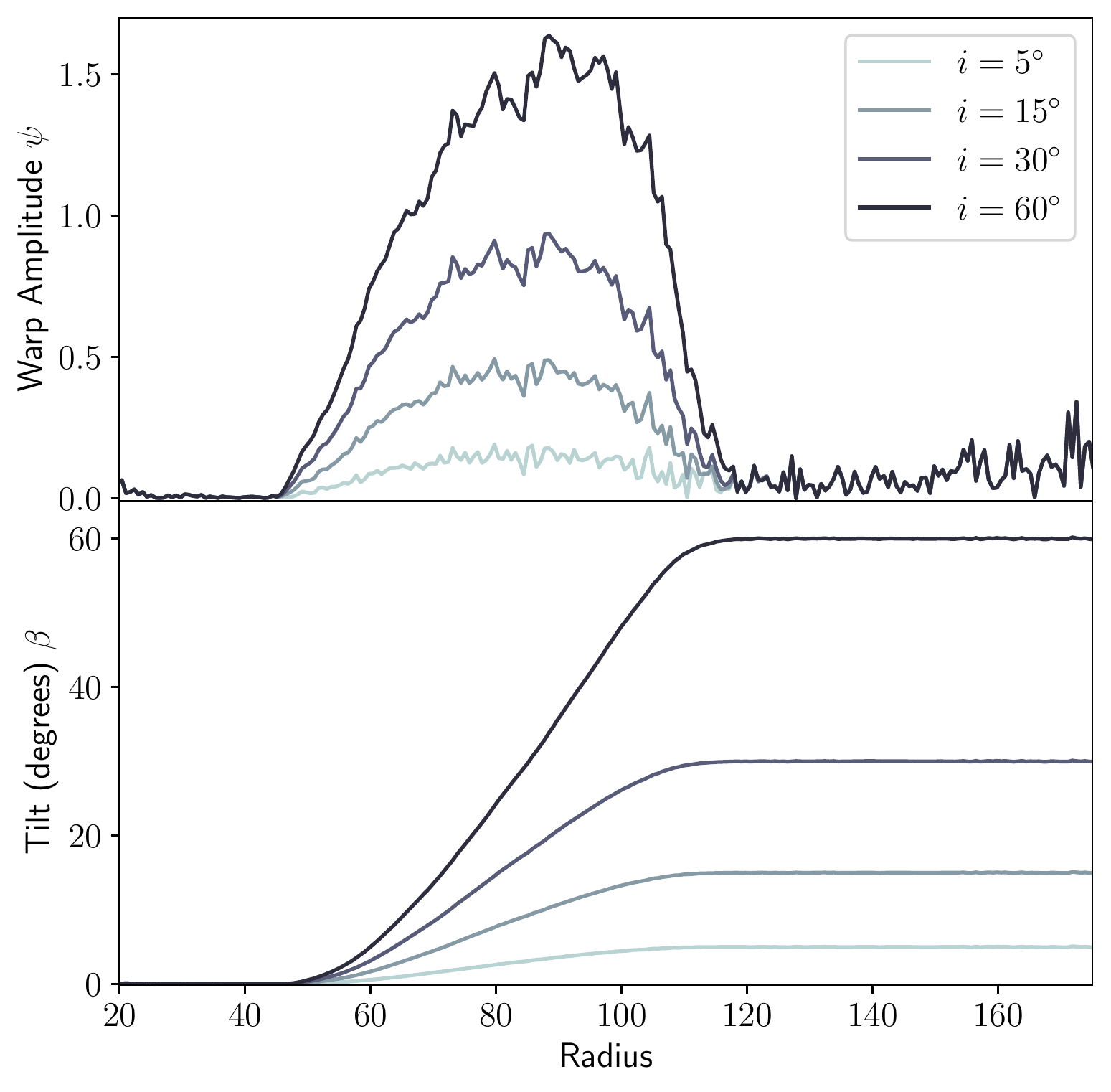}
		\caption{The warp amplitude, $\psi$ (top) and the tilt, $\beta$ (bottom) for the simulations where the initial disc misalignment, $i$ varies from $5^{\circ}$ to $60^{\circ}$. The darker lines represent larger $i$. A larger initial disc misalignment results in a steeper tilt profile, and thus a larger warp amplitude.}
		\label{psiInc}
	\end{center}
\end{figure}

\subsection{Warping the disc}
Before the disc is warped, it is evolved for 10 orbits to allow it to become gravitationally unstable and develop spiral structures. A warped disc can be described by two angles, its tilt $\beta$, and its twist $\gamma$. Using this, the unit angular momentum vector can be written as \citep{1996Pringle}

\begin{equation}
	\vb*{l} = (\cos\gamma \sin\beta,\, \sin\gamma \sin\beta,\, \cos\beta),
\end{equation}
where the disc is considered warped if $\vb*{l}$ varies with radius.
To introduce a warp in our simulation, the position of each particle is rotated such that the unit vector of the angular momentum is given by \citep{2010Lodato}

\begin{gather}
	\begin{aligned}
		l_{x} & = \left\{\begin{array}{ll}
			0                                                                                      & \mathrm{for} \ R < R_{1}         \\
			\frac{A}{2} \left [ 1 + \sin \left ( \pi \frac{R-R_{0}}{R_{2} - R_{1}} \right) \right] & \mathrm{for} \ R_{1} < R < R_{2} \\
			A                                                                                      & \mathrm{for} \ R > R_{2}         \\
		\end{array}\right. \\
		l_{y} & = 0, \\
		l_{z} & = \sqrt{1-l_{x}^{2}},
	\end{aligned}
\end{gather}
where in our fiducial setup the warp is centered at ${R_{0} = 75}$ and extends from $R_{1} = 45$ to $R_{2} = 105$ in code units. The initial misalignment between the outer and inner disc, $i = 30^{\circ}$ defines $A = \sin(i)$. All discs in this work are initially untwisted since $l_y=0$.

Figure \ref{initWarp} also helps to visualise how the warp profile affects the disc's evolution, as also described in equations \ref{eq:pres} and \ref{eq:vr}. For a steeper warp, i.e. a larger warp amplitude $\psi$, a fluid element will experience a larger pressure gradient from adjacent rings as it orbits around the disc. Hence to investigate the importance of the warp amplitude $\psi$ on the disc evolution,  we consider multiple simulations. First, we consider various inclinations where simulations are performed with $i = 5^{\circ}, 15^{\circ}, \ \mathrm{and}\ 60^{\circ}$. By changing only the initial disc misalignment, the only difference to the fiducial setup is the maximum value of $\psi$. Second, the impact of the warp location is also investigated with two additional simulations with the same warp width ($R_{2} - R_{1} = 60$) and initial disc misalignment ($i = 30^{\circ}$), but at $R_{0} = 50$ and $100$.

\section{Results}
\label{sec:Res}

\subsection{Evolution into a gravitationally stable disc}

As the disc in our fiducial simulation evolves, the warp propagates radially both inwards and outwards. The influence of the warp quickly suppresses the spiral structures yielding an axisymmetric gravitationally stable disc. Since the warp is not maintained by external forces, as the disc continues to evolve, the warp smooths out and the disc becomes coplanar. The top panels of Figure \ref{Q_R75i30} shows the surface density of the disc at three snapshots at 0.5, 4, and 14 orbits after the disc is warped. The leftmost panels shows the disc in the early stages when it is still warped. The middle panels show the disc while the warp is dissipating at $t = 4$ orbits. Here, the disc is smooth with no signs of any spiral structures left. The rightmost panels show the disc much later in its evolution after the warp has completely dissipated at $t = 14$ orbits. Due to the constant $\beta_{\mathrm{cool}}$ used to cool the disc, the disc cools back down and becomes gravitationally unstable once more after the warp has dissipated and the disc is back to being flat.

To demonstrate what drives the disc to become gravitationally stable, the azimuthally averaged surface density $\Sigma$, sound speed $c_{s}$, and Toomre $Q$ parameter  \citep{1964Toomre} are plotted in Figure \ref{Q_R75i30azi}, where the darker lines represent later times. The Toomre parameter is given by

\begin{equation}
	Q = \frac{c_{s}\Omega}{\pi G \Sigma}.
\end{equation}
Protoplanetary discs are considered to be gravitationally unstable when spiral structures form. The critical value to form non-axisymmetric instabilities such as spiral arms is when $Q \lesssim 1.7$ \citep{2007Durisen}.  We see that the surface density of the disc is mostly unchanged once the warp is introduced. However, as seen by the increase in $c_{s}$ the influence of the warp heats up the disc by enough to put it in the gravitationally stable regime. During the first couple of orbits when the disc is still warped, Figure \ref{Q_R75i30azi} shows that the disc continues to heat up. As the warp dissipates as the disc  evolves, cooling takes over and the disc starts to cool back down. Due to the simple cooling prescription, the inner regions of the disc begin to cool down first. This is expected in a disc cooled with a constant $\beta_{\mathrm{cool}}$, where the the disc cools faster in the inner disc. Eventually the warp has dissipated such that the disc can cool down enough to reform its spiral structures. The rightmost panels in Figure \ref{Q_R75i30} show the disc at 14 orbits when it is gravitationally unstable again.

% Horizontal version of Figure 5 & 6 as one figure

% \begin{figure*}
% 	\centering
% 	\gridline{\fig{Inc_00025.pdf}{\textwidth}{}}
% 	\gridline{\fig{Inc_00200.pdf}{\textwidth}{}}
% 	\caption{The surface density, $\Sigma$ (left) and divergence of velocity, $\divv$ (right) in code units where the angled brackets represent a density weighted average, showing the evolution of a $0.1M_{\odot}$ disc 0.5 (top) and 4 (bottom) orbits after a warp has been introduced. The subplots are for a disc with initial misalignments of $i =  5^{\circ}, 15^{\circ}, 30^{\circ} \ \mathrm{and} \ 60^{\circ}$. The $\divv$ plots show that the strength of the response to the velocity flow of the disc due to the warp is greater for larger $i$.}
% 	\label{CO}
% \end{figure*}
	
% Figures 5 & 6 vertical format
\begin{figure}
	\centering
	\includegraphics[width=\linewidth]{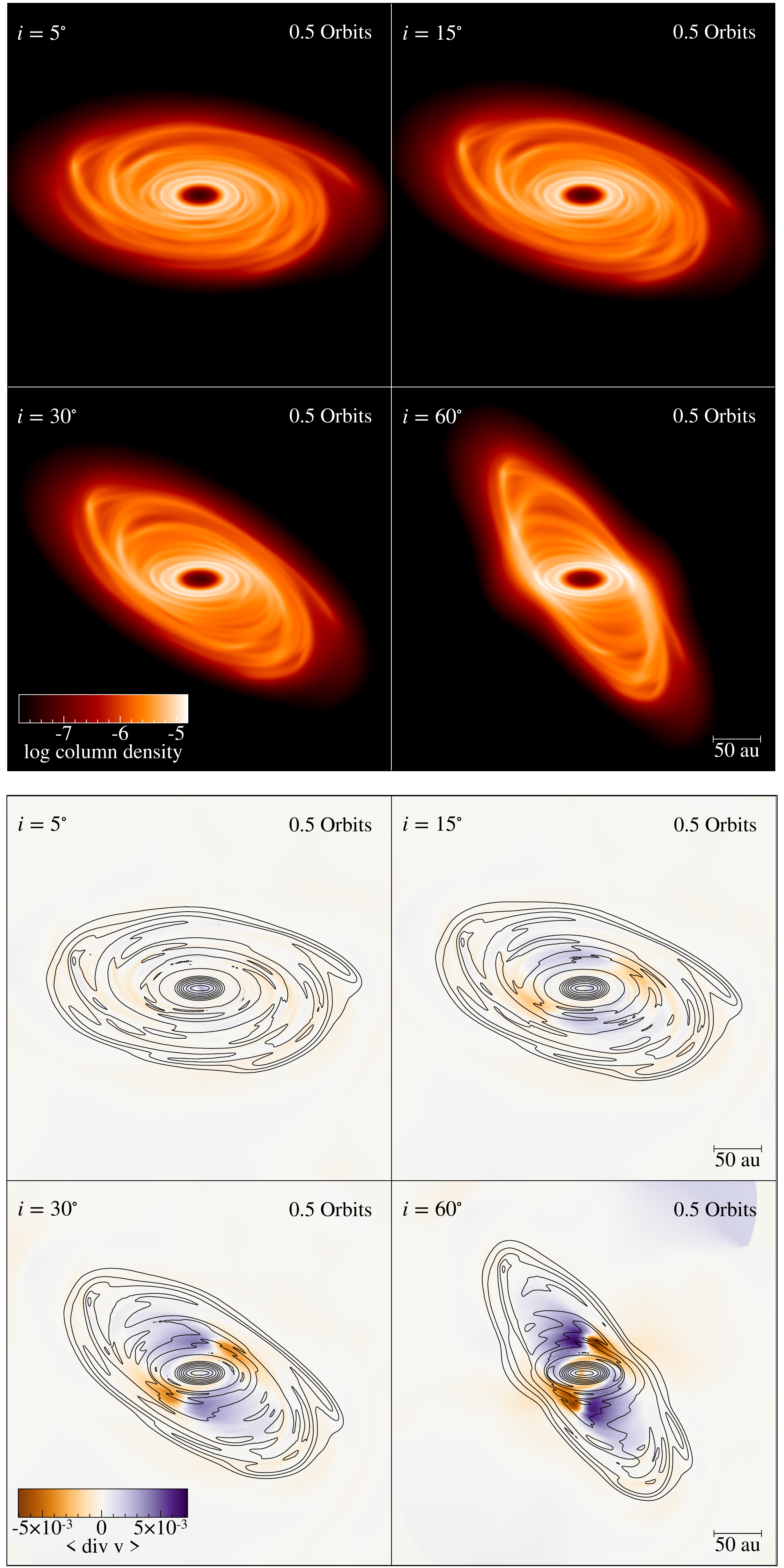}
	\caption{The surface density, $\Sigma$ (top) and divergence of velocity, $\divv$ (bottom) in code units where the angled brackets represent a density weighted average, showing the evolution of a $0.1M_{\odot}$ disc 0.5 orbits after a warp has been introduced. The subplots are for a disc with initial misalignments of $i =  5^{\circ}, 15^{\circ}, 30^{\circ} \ \mathrm{and} \ 60^{\circ}$. The $\divv$ plots show that the strength of the response to the velocity flow of the disc due to the warp is greater for larger $i$.}
	\label{inc}
\end{figure}
\begin{figure}
	\centering
	\includegraphics[width=\linewidth]{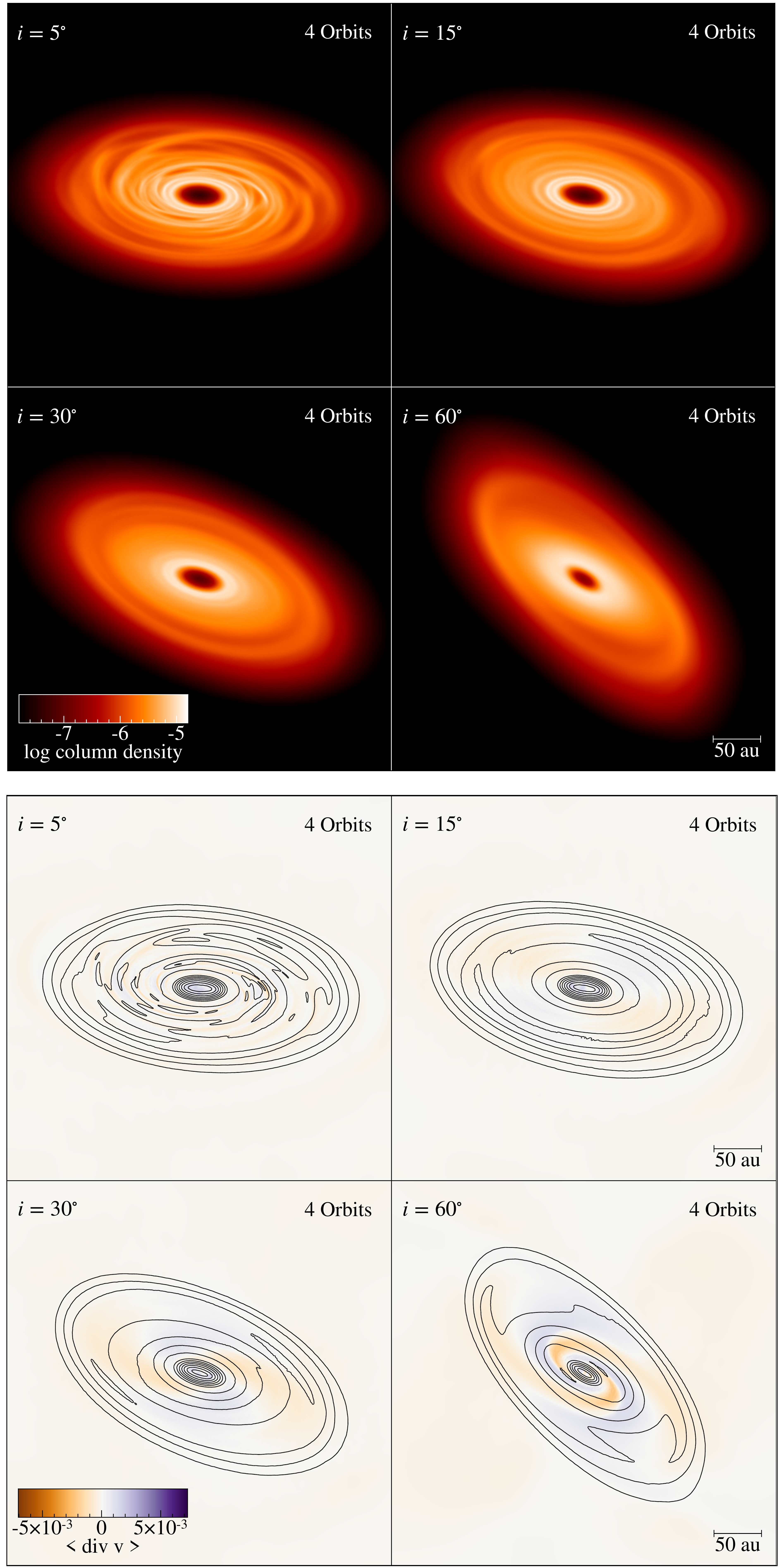}
	\caption{Same as Figure \ref{inc}, but at 4 orbits after the warp has been introduced. Due to the strong response to the velocity flow, the discs in the bottom row ($i = 30^{\circ}$ and $60^{\circ}$) have heated up enough to become gravitationally stable and lose their spiral structures. Whereas for a slight misalignment ($i = 5^{\circ}$), the spiral structures are unaffected as the disc has experienced negligible heating.}
	\label{inc2}
\end{figure}

\begin{figure*}
	\sloppy
	\begin{center}
		\includegraphics[width=\linewidth]{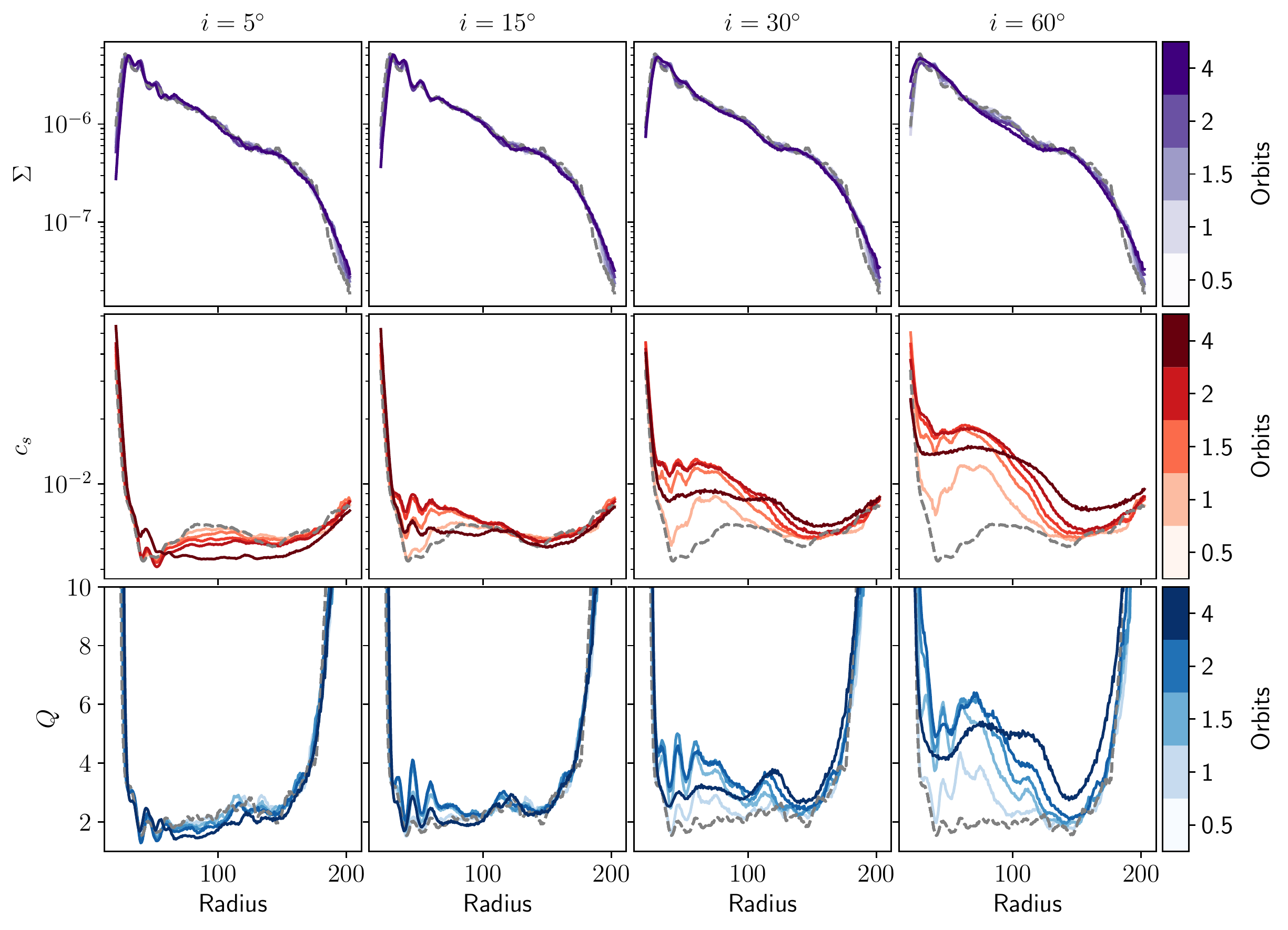}
		\caption{The azimuthally averaged surface density $\Sigma$ (top panels), sound speed $c_{s}$ (middle panels), and Toomre $Q$ parameter (bottom panels) in code units for the discs with initial disc misalignment $i = 5^{\circ}, 15^{\circ}, 30^{\circ}$ and $60^{\circ}$ (from left to right). The darker shades represent later times. The dashed line represents the conditions at $t=0$. The disc only experiences sufficient heating to become gravitational stable when $i = 30^{\circ}$ and $60^{\circ}$.}
		\label{Q_IncEvol}
	\end{center}
\end{figure*}

To determine why the disc initially heats up, we investigate the role of $\PdV$ work. The bottom panel in Figure \ref{Q_R75i30} shows $\divv$, the divergence of the velocity. Recalling from \S\ref{sec:WT} that the warp induces a response in the radial velocity, we examine the consequence of this on the heating in the disc. As the divergence of the velocity directly contributes to the energy (and thus temperature) of the disc, see equation \ref{eq:enrg}, we can use $\divv$ as a proxy for the heating due to $\PdV$ work that is occuring in the disc. In the early stages of evolution at 0.5 orbits, the disc is still warped. As mentioned in \S\ref{sec:WT}, strong radial pressure gradient are induced due to adjacent annuli of gas varying with vertical height $z$. This triggers a response in the induced radial velocity of the disc which heats it up as seen by the large magnitude of $\divv$ in the bottom left panel of Figure \ref{Q_R75i30}. As the disc evolves, the warp dissipates as there is no external torque sustaining the warp. At 4 orbits, the disc is nearly flat and hence there is little variance in the pressure gradient. Thus, the magnitude of $\divv$ has greatly decreased, as expected, leading to less heating from PdV work.  At 14 orbits, the only contributions to $\PdV$ heating is due to the gravitational instabilities in the disc, which is much weaker compared to the $\PdV$ work due to the warp.

% \begin{figure*}
% 	\sloppy
% 	\begin{center}
% 		\includegraphics[width=\linewidth]{Q_evol_inc.pdf}
% 		\caption{The azimuthally averaged surface density $\Sigma$ (top panels), sound speed $c_{s}$ (middle panels), and Toomre $Q$ parameter (bottom panels) in code units for the discs with initial disc misalignment $i = 5^{\circ}, 15^{\circ}, 30^{\circ}$ and $60^{\circ}$ (from left to right). The darker shades represent later times. The dashed line represents the conditions at $t=0$. The disc only experiences sufficient heating to become gravitational stable when $i = 30^{\circ}$ and $60^{\circ}$.}
% 		\label{Q_IncEvol}
% 	\end{center}
% \end{figure*}

\subsection{Impact of warp inclination}
\label{sec:inc}

Figure \ref{psiInc} shows the initial warp profile for three additional simulations, each with a different initial disc misalignment of $i = 5^{\circ}, 15^{\circ}, 60^{\circ}$ to compare with the fiducial simulation with $i = 30^{\circ}$. Recalling eq \ref{eq:pres}, the pressure gradient induced by the warp depends on the pressure $p$, disc scale height $H$, and warp amplitude $\psi$. By fixing $R_{0}$ and $R_{2} - R_{1}$, the location and extent of the warp respectively, $p$ and $H$ are unchanged in the warped regions for all four discs. Thus, only changing the maximum value of $\psi$ allows for easier comparison of how the warp affects the evolution of the disc.

Figures \ref{inc} and \ref{inc2} show the surface density and the divergence of the velocity at 0.5 and 4 orbits respectively after the warp has been introduced. From the plots of $\divv$ in Fig \ref{inc} it is clear that, all else being equal, the discs with the stronger warp profiles trigger greater responses in the induced radial velocity of the disc. This is easiest seen comparing the two extremes where $i = 5^{\circ}$ and $i = 60^{\circ}$ (top left and bottom right sub panels). For the disc where $i = 5^{\circ}$, there is little change to the $\divv$ of the disc. Whereas with $i = 60^{\circ}$, there is a large impact on $\divv$ throughout most of the disc. Fig \ref{inc2} shows the consequence of the warp on the disc structure. It can be seen that the discs with stronger misalignments ($i = 30^{\circ}$ and $i = 60^{\circ}$ in particular), which trigger stronger responses in the induced radial velocity of the disc, experience greater heating resulting in the discs becoming gravitationally stable with no signs of spiral structures after a few orbits.

This is reflected in Figure \ref{Q_IncEvol} which shows the azimuthally averaged surface density $\Sigma$ (top panels), sound speed $c_{s}$ (middle panels), and Toomre $Q$ parameter (bottom panels). In all cases, the surface density is mostly unchanged as with the fiducial case. The change in the sound speed on the other hand has a strong dependence on the initial disc misalignment. In general, a larger initial disc misalignment, and hence a larger warp amplitude $\psi$, results in more heating. However, simply having a misaligned disc does not necessarily result in the disc becoming gravitationally stable. With a slight misalignment of $i = 5^{\circ}$, since there is very little change to the velocity flow of the disc, the contribution to the $\PdV$ work due to the warp is negligible and the disc simply continues to cool as expected due to the constant $\beta_{\mathrm{cool}}$ used in these simulations. Although there is some heating in the $i = 15^{\circ}$ case, it is not enough to push the disc into the gravitationally stable regime as it quickly cools back down to pre-warp levels. It is only in the $i = 30^{\circ}$ and $60^{\circ}$ cases that there is enough heating to result in the disc becoming completely gravitationally stable.

\begin{figure}
	\begin{center}
		\includegraphics[width=\linewidth]{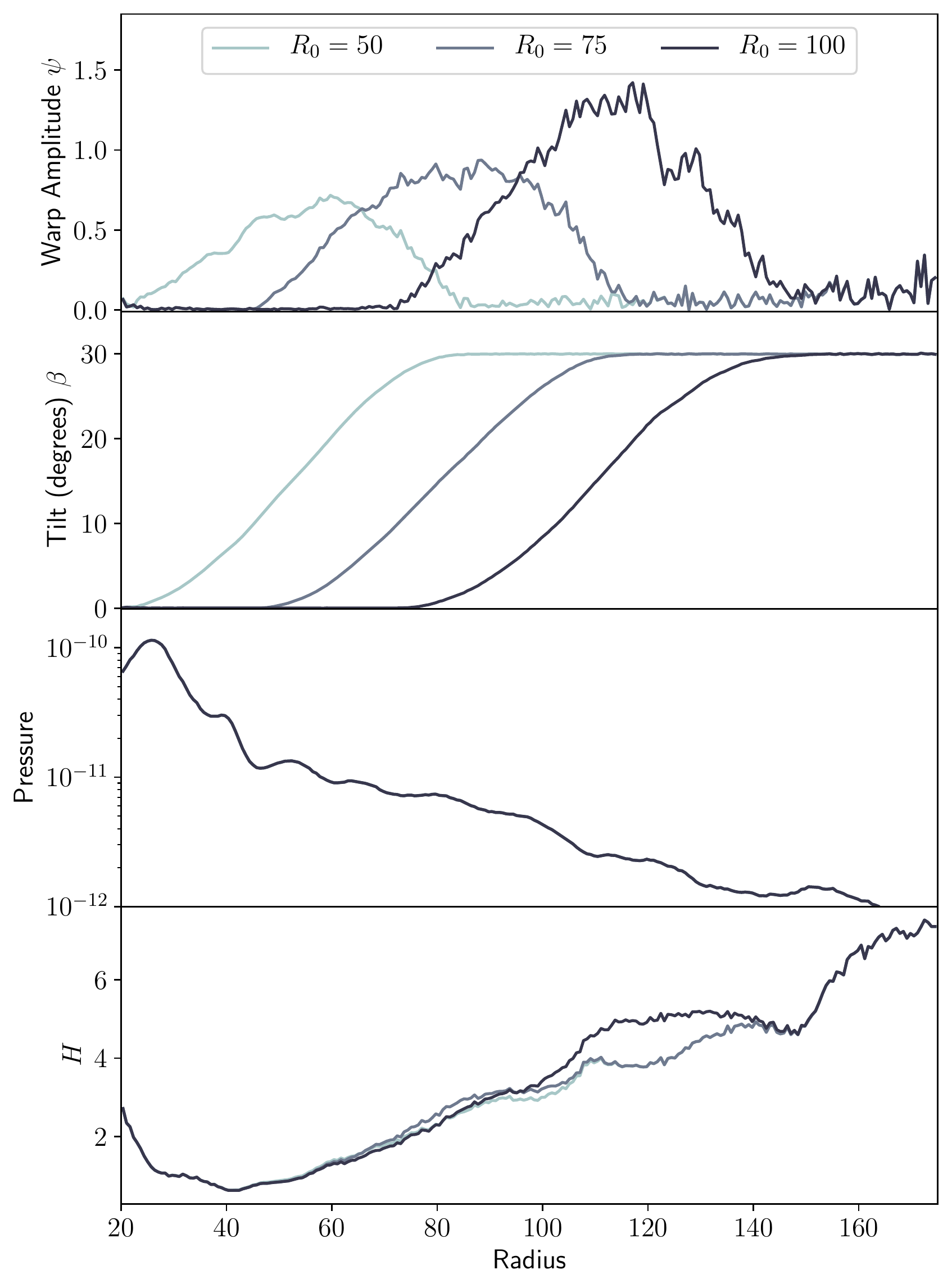}
		\caption{The warp amplitude, $\psi$ (first panel) and the tilt, $\beta$ (second panel) for the simulations where the warp location, $R_{0} = 50, 75$, and $100$. The darker lines represent larger $R_{0}$. The maximum value of $\psi$ increases at higher $R_{0}$. The pressure and disc scale height in code units are shown in the third and fourth panels respectively. The pressure profile is the same for all simulations, hence the lines are coincident.}
		\label{psiRwarp}
	\end{center}
\end{figure}

\subsection{Impact of warp location}

Unlike in \S\ref{sec:inc} where varying the initial disc misalignment $i$, only changes the warp profile, varying the warp location $R_{0}$, adds additional complexities to the evolution of a warped disc. Figure \ref{psiRwarp} shows the warp profiles (first two panels), the initial pressure $p$ (third panel), and disc scale height $H$ (bottom panel), for the simulations with $R_{0} = 50, 75$ and $100$. The bottom two panels are important to understanding the pressure gradients induced by the warp, recalling eq \ref{eq:pres}.

\begin{figure}
	\begin{center}
		\includegraphics[width=\linewidth]{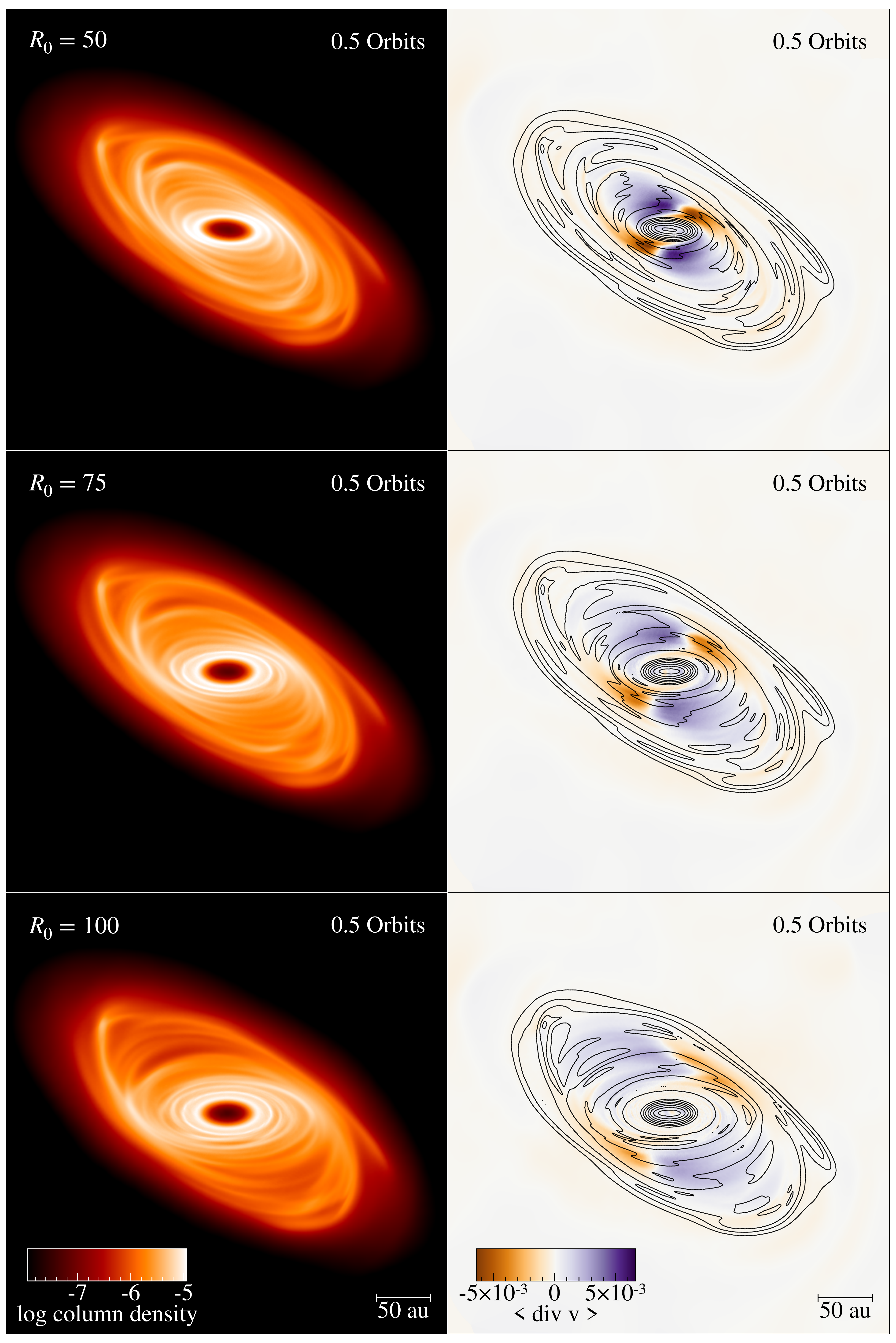}
		\caption{The surface density, $\Sigma$ (left) and divergence of velocity, $\divv$ (right) in code units showing the evolution of a $0.1M_{\odot}$ disc 0.5 orbits after a warp has been introduced. The subplots are for a disc with the warp located at $R_0 =  50, 75\ \mathrm{and} \ 100$. Although the warp amplitude $\psi$ increases with higher $R_0$, the magnitude of $\divv$ is largest at lower $R_0$, which is at odds with the results in Figure \ref{inc}. However, once the differences in pressure $p$, and disc scale height $H$ at different $R_0$ are considered, this discrepancy is reconciled. Higher pressures and and a smaller scale height at lower $R_0$ counter the lower warp amplitude giving rise to greater response in the induced radial velocity.}
		\label{Rwarp25}
	\end{center}
\end{figure}

Since the warp amplitude, $\psi$ is larger for increasing $R_{0}$, based on the results in \S\ref{sec:inc} we might naïvely expect the magnitude of $\divv$ to be largest for the disc warped at $R_{0} = 100$. However, from Figure \ref{Rwarp25} this is clearly not the case. Instead, the largest change to $\divv$ is for the disc warped at $R_{0} = 50$ which has the smallest $\psi$. To reconcile this discrepancy, we have to also consider the pressure and scale height at each $R_{0}$. From Fig \ref{psiRwarp}, the pressure increases while the disc scale height decreases for smaller $R_{0}$. Both of these work towards increasing the pressure gradients induced by the warp (see equation \ref{eq:pres}), thus triggering a stronger response in $\divv$ for smaller $R_{0}$, which is consistent with our results.

\begin{figure}
	\begin{center}
		\includegraphics[width=\linewidth]{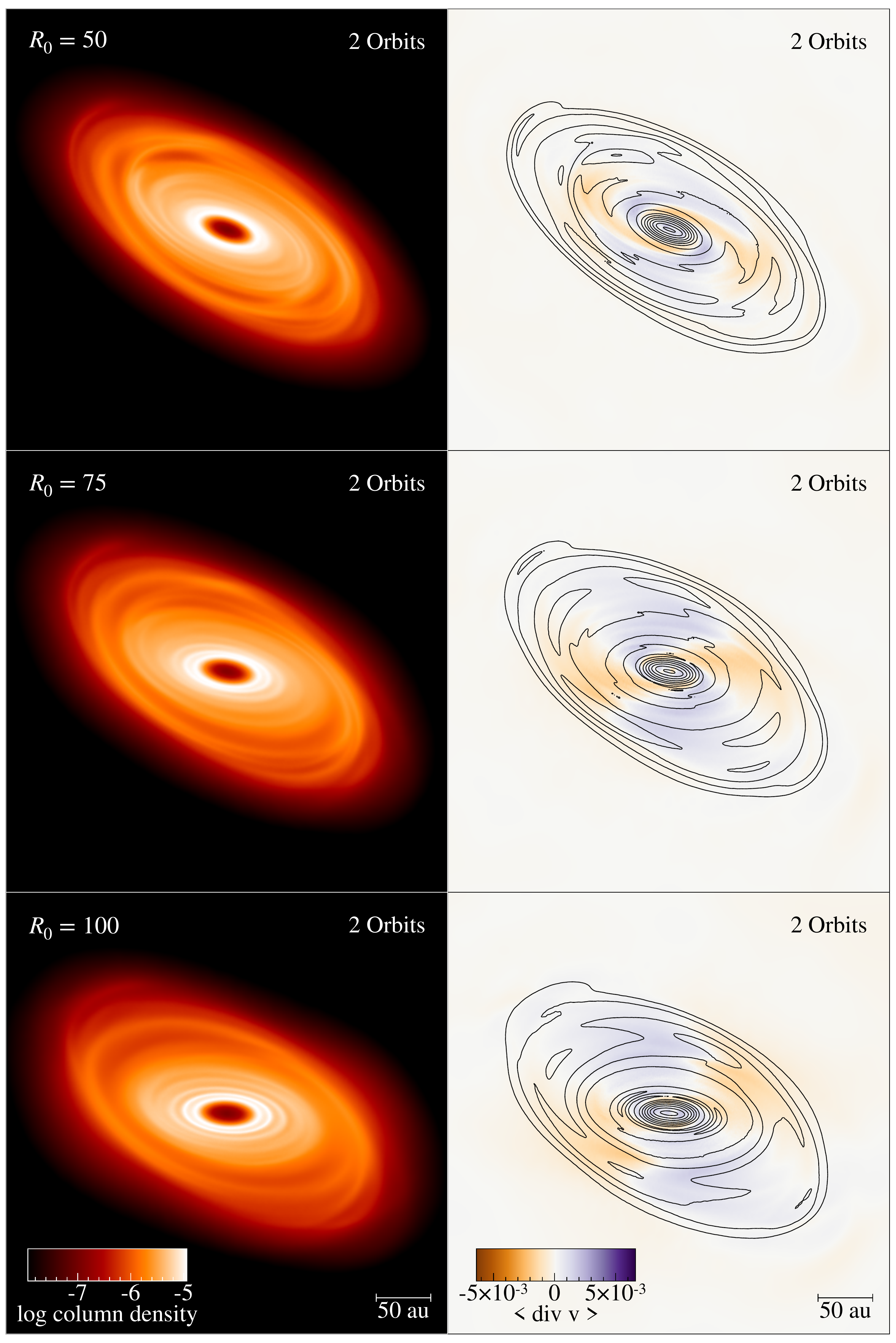}
		\caption{Same as Figure \ref{Rwarp25} but at 2 orbits after the warp has been introduced. Despite the magnitude of $\divv$ being initially larger at lower $R_0$, the impact to the spiral structures is greater at larger $R_0$. This result initially appears at odds with Figure \ref{inc2} where the spiral structures are most suppressed with increasing magnitudes of $\divv$. However consideration of the cooling and realignment timescales reconciles this discrepancy. Both timescales are shorter at smaller $R_0$. Hence, we expect the magnitude of $\divv$, and hence the amount of heating, to decrease faster for discs warped at lower $R_0$, which can be seen by comparing  with the right panels of Figure \ref{Rwarp25} to see how much the magnitude of $\divv$ has changed by. Therefore, the results here are consistent with our expectations of spiral structures being most impacted at larger $R_0$ since the heating due to the warp is prolonged and more efficient at larger $R_0$.}
		\label{Rwarp100}
	\end{center}
\end{figure}

\begin{figure*}
	\begin{center}
		\includegraphics[width=\linewidth]{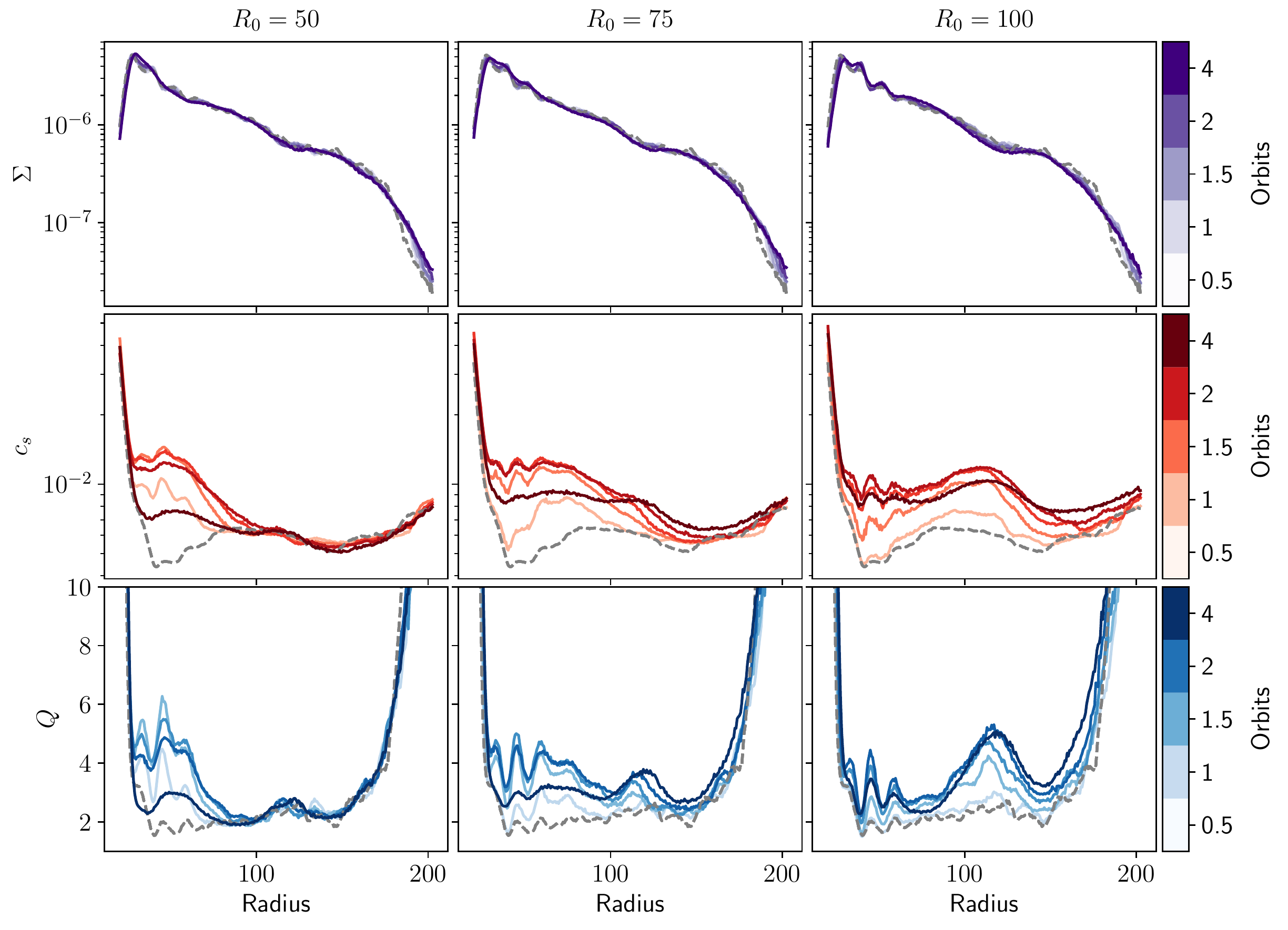}
		\caption{The azimuthally averaged surface density $\Sigma$ (top panels), sound speed $c_{s}$ (middle panels), and Toomre $Q$ parameter (bottom panels) in code units for the discs with the warp located at $R_{0} = 50, 75$, and $100$ (from left to right). The darker shades represent later times. The dashed line represents the conditions at $t=0$. The disc experiences greater heating, and thus is more impacted, at larger $R_0$ due to longer cooling and realignment timescales at larger radii.}
		\label{Q_RwarpEvol}
	\end{center}
\end{figure*}

Based on the results in \S\ref{sec:inc} we should expect that the spiral structures to be most weakened for the discs warped at smaller $R_{0}$, since they have the largest change in $\divv$. However, from Figure \ref{Rwarp100}, this is clearly not the case. Instead, the impact to the disc structure is greater at larger $R_{0}$. To understand why, we have to now consider the timescale at which warp propagation is damped, $t_{\mathrm{damp}} = (\alpha \Omega)^{-1}$ \citep{2000Lubow}, and the cooling timescale, $t_{\mathrm{cool}} = \beta_{\mathrm{cool}}\Omega^{-1}$ \citep{2001Gammie}. Both of these timescales are longer at larger $R_{0}$. A longer $t_{\mathrm{damp}}$ results in the disc taking longer to realign. Since the disc is in a warped state for an increased duration, the heating due to the warp is prolonged. Comparing the right panels of Figs \ref{Rwarp25} and \ref{Rwarp100} shows that the magnitude of $\divv$ has decreased the most for $R_{0} = 50$. Whereas for $R_{0} = 100$ the magnitude of $\divv$ has not changed too much. This is consistent with the disc realigning much faster for a warp at $R_{0} = 50$. Since $t_{\mathrm{cool}}$ is also shorter at smaller $R_{0}$, cooling is faster in the inner regions of the disc due to the simple cooling model used. Hence, the heating due to the warp is less efficient for smaller $R_0$.

Figure \ref{Q_RwarpEvol} shows the azimuthally averaged surface density $\Sigma$ (top panels), sound speed $c_{s}$ (middle panels), and Toomre $Q$ parameter (bottom panels) for these simulations. The impact of the shorter cooling and realignment timescales at smaller $R_0$ is more apparent in this figure. The disc with $R_0 = 50$ cools the quickest. Additionally, since the disc also realigns the fastest for $R_0 = 50$, the heating due to the warp does not last long enough to spread throughout the disc. This is in contrast to the disc with $R_0 = 100$ which has experienced the most heating globally due to a longer realignment timescale. In general, we find that the spiral structures are more affected by a warp located further out which results in the disc being heated up for a longer duration.

\section{Discussion}
\label{sec:disc}

\subsection{Disc cooling \& long term evolution}

The cooling parameter, $\beta_{\mathrm{cool}}$ is constant both spatially and temporally. Although commonly used to model self-gravitating discs, comparison with radiative transfer models show that $\beta_{\mathrm{cool}}$ should vary in both time and space \citep{2018Mercer}. For the purposes of this work, a constant $\beta_{\mathrm{cool}}$ is justified as we are primarily interested in how the evolution of the disc structure responds to a warp. Since $\PdV$ heating dominates over cooling while the disc is warped, we expect the qualitative results to hold; a warp acts to move a self-gravitating disc towards stability. 

However, the long term evolution of the disc will be better understood when taking into account the non-coplanar geometry with more sophisticated disc thermodynamics. In all simulations where the disc becomes gravitationally stable due to the warp, the disc eventually cools back down to become gravitationally unstable and recover its spirals throughout the disc. This is simply a result of the cooling model which is time-independent and thus does not take into account the structural evolution of the disc. Additionally, the shorter cooling time in the inner disc naturally results in spiral structures forming more quickly in the inner regions, contrary to expectation of realistic self-gravitating discs \citep{2005Rafikov, 2009Stamatellos, 2009Rice, 2009Clarke}. Hence, more realistic thermodynamics are needed to better understand the longer term evolution of the suppression of spiral structures due to the warp.

% Spiral structures forming in the inner parts earliest is not expected in realistic self-gravitating discs where it should only be gravitationally unstable in the outer regions \citep{2005Rafikov, 2009Stamatellos, 2009Rice, 2009Clarke}. 

\subsection{Formation of rings \& gaps}

An interesting result which could have important implications on observations is the formation of rings and gaps from the evolution of a warped gravitationally unstable disc. We find tentative evidence of ring and gap structure, most easily seen in the bottom row of Figures \ref{inc2} ($i = 30^{\circ}$ and $60^{\circ}$) and \ref{Rwarp100} ($R_0 = 100$). The evolution of the warp amplitude which has a continuous decline in magnitude combined with the surface density profiles suggests that the gap is not a result of disc breaking. We also do not expect the disc to break due to the high $\alpha$ viscosities \citep{2018Dogan} associated with gravitationally unstable discs generally and in our simulations ($\alpha \sim 0.1$). Our results instead suggest that the spiral structures initially present in our disc evolved into ring and gap structure as a result of the warp propagation. However, the morphology of the spiral structures depend on the disc thermodynamics. Thus, more detailed simulations are required to determine if warps in gravitationally unstable discs can explain some of the sub-structures seen in observations. This may explain features in existing discs such as Elias 2-27, which contains a gap, spiral structures and evidence of a warp \citep{2021Paneque}. If the warp in Elias 2-27 is located in the inner parts of the disc, the impact of the warp could be localised to the inner disc allowing the spiral structures in the outer regions of the disc to survive.

\section{Conclusion}
\label{sec:conc}

We perform 3D SPH simulations to investigate the impact of a warp on the structure of a gravitationally unstable disc. Our work shows that if the warp is strong enough, it can suppress spiral structures due to gravitational instability. This is due to the oscillating radial pressure gradient induced by the warp which triggers a response in the velocity flow of the disc. This causes the disc to heat up and become gravitationally stable. In some cases, the disc evolves to form ring \& gap structure, which could have important observational implications. 

We find that that the structure is more impacted for warps with larger initial disc misalignments and for warps located further out. In the former case, a larger initial misalignment simply results in larger pressure gradients due to a larger warp amplitude, and thus a stronger response in the velocity flow of the disc leading to increased heating. The latter case is more complicated due to physical properties of the disc varying with radius, but a warp located further out has a greater impact on the spiral structure due to a longer realignment timescale, which results is prolonged heating. Finally, we note that a more detailed treatment of the disc thermodynamics is required to understand the long term evolution of the tentative ring and gap structure we identify in some of our simulations.

\section*{acknowledgments}
%\acknowledgements

We thank Giuseppe Lodato for useful discussions and the anonymous referee for their comments that benefited this work. SR acknowledges support from the Royal Society Enhancement Award. FM acknowledges support from the Royal Society Dorothy Hodgkin Fellowship. RN acknowledges support from UKRI/EPSRC through a Stephen Hawking Fellowship (EP/T017287/1). This work was performed using Orac, the HPC cluster at the University of Warwick.

\software{Matplotlib \citep{Hunter:2007},
	numpy \citep{5725236},
	pandas \citep{reback2020pandas,mckinney-proc-scipy-2010},
	\textsc{phantom} \citep{2018Price},
	\textsc{splash} \citep{2007Price}
}

\bibliography{HidingGI_Warps.bib}{}
\bibliographystyle{aasjournal}

\end{document}